\documentclass{aastex}
\usepackage{graphicx}
\usepackage{subfigure}
\usepackage{natbib}
\usepackage{ifpdf}

\shorttitle{Intrinsic Metallicity Dispersions in Dwarf Galaxies and Star Clusters}
\shortauthors{Leaman}

\begin{document}
\title{Insights into Pre-Enrichment of Star Clusters and Self-Enrichment of Dwarf Galaxies from their Intrinsic Metallicity Dispersions}
\author{Ryan Leaman$^{1}$}
\affil{$^{1}$University of Victoria, Canada}
\email{rleaman@uvic.ca}

\begin{abstract}
Star clusters are known to have smaller intrinsic metallicity spreads than dwarf galaxies due to their shorter star formation timescales.  Here we use individual spectroscopic [Fe/H] measurements of stars in 19 Local Group dwarf galaxies, 13 Galactic open clusters, and 49 globular clusters to show that star cluster and dwarf galaxy linear metallicity distributions are binomial in form, with all objects showing strong correlations between their mean linear metallicity $\bar{Z}$ and intrinsic spread in metallicity $\sigma(Z)^{2}$.
A plot of $\sigma(Z)^{2}$ versus $\bar{Z}$ shows that the correlated relationships are offset for the dwarf galaxies from the star clusters.  The common binomial nature of these linear metallicity distributions can be explained with a simple inhomogeneous chemical evolution model (e.g., \cite{Oey00}), where the star cluster and dwarf galaxy behaviour in the $\sigma(Z)^{2}-\bar{Z}$ diagram is reproduced in terms of the number of enrichment events, covering fraction, and intrinsic size of the enriched regions.  The inhomogeneity of the self-enrichment sets the slope for the observed dwarf galaxy $\sigma(Z)^{2}-\bar{Z}$ correlation. The offset of the star cluster sequence from that of the dwarf galaxies is due to pre-enrichment, and the slope of the star cluster sequence represents the remnant signature of the self-enriched history of their host galaxies.  The offset can be used to separate star clusters from dwarf galaxies without a priori knowledge of their luminosity or dynamical mass.  The application of the inhomogeneous model to the $\sigma(Z)^{2}-\bar{Z}$ relationship provides a numerical formalism to connect the self-enrichment and pre-enrichment between star clusters and dwarf galaxies using physically motivated chemical enrichment parameters. Therefore we suggest that the $\sigma(Z)^{2}-\bar{Z}$ relationship can provide insight into what drives the efficiency of star formation and chemical evolution in galaxies, and is an important prediction for galaxy simulation models to reproduce.
\end{abstract}


\section{Introduction}
Correlations between the average metallicity and the mass or luminosity of a galaxy (e.g., \citealt{Lequeux79,Skillman89b}), the mass-metallicity relation (MZR), offers a window into how chemical evolution in galaxies proceeded over cosmic times.  Numerous studies have explored the interplay between chemical enrichment produced via SNe versus the efficiency of outflows - and in recent years gas inflows \citep{Keres05,Brooks07,Brooks09}.  However, a complete picture for how the MZR came to be at redshift zero, and evolved before that, is still missing at this time.   
The star formation (SF) and metal retention efficiencies as a function of galaxy mass are also 
needed to understand the distribution of baryons in dark matter halos.  For example,
to what degree is SF modulated by metallicity dependent H$_{2}$ formation \citep{Kuhlen11}, 
heating from stellar and SN feedback \citep{Governato10}, or environment \citep{Mayer06}.
It is also not clear why galaxies with masses less than $\sim 10^{10} \rm{M_{\odot}}$ have 
such low baryon fractions relative to Milky Way (MW) sized galaxies (e.g., \citealt{Guo10}).

Along with the mean metallicity of a host system, comparisons of the intrinsic spread in metallicity for different objects may help constrain the timescale for chemical enrichment in galaxies.  There is increasing observational evidence for inhomogeneous mixing in dwarf galaxies (e.g., \citealt{Koch08}; \citealt{Venn12}), or alternatively hierarchical merging of minihalos \citep{Salvadori09,Salvadori12}, both of which can be traced by the dispersion in elemental abundances.  

\cite{Kirby11a} and \cite{Norris10} each looked at the spread in [Fe/H] versus host galaxy luminosity (L) amongst Local Group dwarf spheroidals (dSphs) and ultra faint dwarfs (UFDs), with both studies finding a clear anticorrelation.  Leaman et al. (2012b) included higher luminosity dwarf irregular (dIrr) systems and found that the spread in [Fe/H] tended to plateau above $L = 10^{5} L_{\odot}$.  Alternatively, \cite{Kirby11a} showed that a strong correlation between spread in linear metal fraction, $\sigma(Z)$ and luminosity exists, with much tighter scatter than in the $\sigma([Fe/H]) - L$ plane.  
It is currently unclear whether these trends in metallicity spread are due to SF duration or enrichment efficiencies.  Nevertheless such correlations may be useful to infer the relative formation and enrichment timescales for the UFDs and dSphs (i.e.,\citealt{Willman11}).  In fact, the
similarities in metallicity spreads between the UFDs and the more luminous dSphs is one item used to differentiate the UFDs from simple star clusters \citep{Willman12}.

 Both star clusters and dwarf galaxies (DGs) have resolved spectroscopic observations of many stars, however they are treated independently in chemical evolution studies.  Specialized n-body+SPH simulations exist for both dwarf galaxies and star clusters (e.g., \citealt{Governato10,Nakasato00}), however they are not general enough, with sufficiently high resolution, to analyze both objects in a single simulation (although see \citealt{Griffen10}).
Therefore, comparative chemical analyses of star clusters and dwarf galaxies have been typically done through analytic descriptions.  Chemical evolution models that relax some of the homogeneity and mixing assumptions from the simple one-zone closed box models have also been developed, and these are able to parameterize physical processes (such as stellar feedback and outflow) in low mass systems (e.g., \citealt{vandenhoek97}; \citealt{Argast00}; \citealt{Karlsson05}).   These chemical evolution models also incorporate stochastic sampling effects, allowing for detailed predictions in the elemental abundance distributions.

Specialized inhomogeneous mixing models can generate simulated metallicity distribution functions (MDFs), however comparisons of these to observed MDFs are very sensitive to biases in the sampling, radial coverage, and measurement methods.   In this paper, we use an inhomogeneous stochastic chemical evolution model to compare only the intrinsic spread and mean metallicity to observationally well sampled star clusters and dwarf galaxies. This analysis does not require an estimate of the host galaxy luminosity (a quantity which may be strongly affected by tidal stripping), and exploits specific statistical properties of the MDFs which we demonstrate are universal.  The scaling relations developed here show the link between the formation of star clusters and chemical evolution of dwarf galaxies, through a simple binomial distribution in metal enrichment events.

\section[]{Data Sources}
The data for this paper come from published spectroscopic studies of individual stars in Local Group dwarf galaxies, open clusters, and globular clusters in the MW.  [Fe/H] values and errors were used to compute metallicity distributions and intrinsic dispersions uniformly.  The metallicity distributions of the star clusters are known to exhibit small spreads in [Fe/H] \citep{CG09} and were initially included only to offer an interesting benchmark to compare to the dwarf galaxy distributions and dispersions. 

In assembling the data, an effort was made to use homogeneous studies of several objects observed with a single instrument setup and analysis technique.  However in cases where a choice of studies were available, data were selected where: 1) spectral resolving power was the greatest, 2) the number of stars in the study was large, 3) in the case of Calcium triplet (CaT) based [Fe/H] measurements, the equivalent width data of individual stars was available allowing us to transform these to [Fe/H] using the improved non-linear calibration by \cite{Starkenburg10}, and 4) the spatial completeness was high.  
For some systems, such as the UFDs, the small number of stars is due to the intrinsic low luminosity of the system.
These small number statistics must be kept in mind as the sample distributions may deviate from the true system distributions, especially in the case of spatial incompleteness and radial population gradients (c.f., Leaman et al., 2012b).  
We restrict ourselves to spectroscopic abundances to avoid the age-metallicity degeneracies which plague photometric metallicities.    Table 1 lists the objects considered in this Paper, along with the original literature study, number of stars with spectroscopic metallicity measurements, resolving power of the spectrograph used, and [Fe/H] measuring method - CaT, or medium resolution/high resolution spectroscopy (MRS/HRS).  

Spectroscopic measurements of individual stars are used to derive [Fe/H] abundances by comparing synthetic spectra to observed Fe lines, by measuring the equivalent width of Fe I and FeII lines, or relating the Calcium Triplet equivalent width to [Fe/H] using empirical relations (i.e., \citealt{ADC91}; \citealt{Starkenburg10}).
The individual [Fe/H] measurements are taken directly as listed in each of the literature studies, with a few exceptions:  
In the case of high resolution abundances, the [Fe/H] measurements were taken as the weighted average of the [Fe\textsc{I}/H] and [Fe\textsc{II}/H] values, however their variation was typically less than $5\%$. 
In surveys using the CaT, [Fe/H] values were recomputed using the two strongest CaT lines and the non-linear relations 
of \cite{Starkenburg10}.  This recalibration ensures that the non-linear bias at faint magnitudes and low metallicities 
is removed.
Another exception is the open cluster survey by \cite{Warren09} and \cite{Cole04} where the V or K band magnitude calibrations (created in those papers) were used.  The near solar metallicity of the open clusters means that [Fe/H] biases are negligible, unlike in the lower metallicity globular clusters or dwarf galaxies.  
Finally, the high resolution studies may have used different solar abundances for their zero points (e.g., log$_{\epsilon_{Fe}}$ = 7.52 from \cite{GS98}, or 7.45 from \citealt{Asplund09}), however the differences are very small, having no significant effect on the results presented in this paper.

\section{Metallicity Distributions}
With a relatively homogeneous set of [Fe/H] values for the dwarf galaxies, open clusters, and globular clusters, metallicity distribution functions can be computed for all objects.  This was first done in [Fe/H] space, and subsequently, each individual stellar [Fe/H] measurement was transformed into a linear metal fraction Z, assuming $Z = Z_{\odot}10^{[Fe/H]}$.  This conversion has two sources of uncertainty; the first due to the exclusion of contributions from $\alpha$-elements to the total metal fraction. 
At low metallicities the $\alpha$ element contribution is larger, but is roughly constant with metallicity (e.g., \citealt{Venn04}).  
Variations in [$\alpha$/Fe] are noticeable ($\leq 0.4$ dex) at intermediate metallicities, particularly in dwarf galaxies 
(e.g., Tolstoy et al. 2009), however inclusion of [$\alpha$/Fe] in each galaxy is difficult and beyond the scope of this paper.   
Therefore, while a true estimate of the total linear metal fraction is not encompassed in the $Z$ value, 
we use $Z$ to track the contribution of Fe as a simple ``linear [Fe/H]'' or ``linear metallicity'' in order to study the metal dispersions in a linear scale. 
The second source of uncertainty is in the solar value of $Z_{\odot}$, which may be as high as $50\%$ \citep{vandenberg07}.  The adopted $Z_{\odot}$ values from the literature represents a small offset that may contribute to the scatter 
in the relations discussed in $\S4$.

\begin{figure}
\begin{center}
\ifpdf
\includegraphics[width=0.99\textwidth]{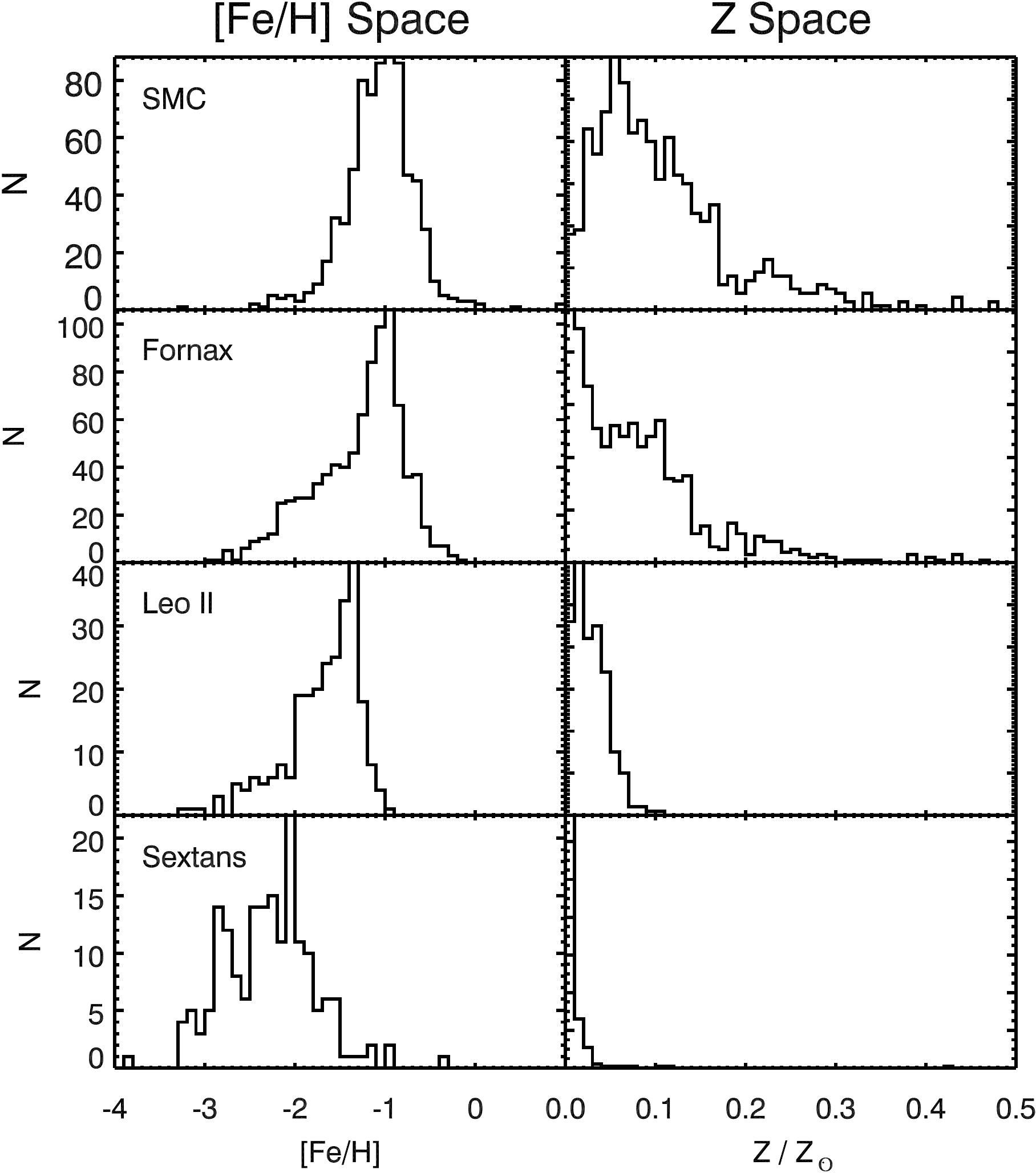}
\else
\includegraphics[width=0.99\textwidth,angle=-90]{distcompt.eps}
\fi
\caption{Sample metallicity distributions for four dwarf galaxies.  Left panels show raw [Fe/H] distribution, while the right panels shows the distribution of linear metallicities, Z.   We note that the spread in the linear metallicity distribution is correlated with the mean linear metallicity.}
\label{fig:distcomp}
\end{center}
\end{figure}

The metallicity distribution functions in [Fe/H] and $Z$ are shown in the left and middle panels of Figure \ref{fig:distcomp} for four representative dwarf galaxies.  While the [Fe/H] distributions are relatively Gaussian, the linear metallicities 
follow a Poisson, or binomial, distribution (especially noticeable in the lower metallicity systems such as Sextans).  
Comparing the linear metallicity spread $\sigma(Z)$ to metallicity Z or other galaxy properties 
must be done with care when the distributions are Poisson or binomial, as the variance of such distributions is 
proportional to the mean metallicity of the object, and physical comparisons between objects may be difficult to 
disentangle from the statistical properties of the distribution itself.  
Thus, it should be kept in mind that spatial biases in a sample will influence the spread in [Fe/H] and Z differently.  
For example, a galaxy with a steep negative metallicity gradient which is sampled preferentially in the inner regions will return a sample mean which is obviously higher than the total galaxy for both [Fe/H] and Z, however the intrinsic spread in Z will be decreased less than the spread in [Fe/H] due to the tail of the Z distributions being in the metal-rich regime.

\subsection{Quantifying the Statistical Form of the $Z$ Distributions}
There is a strong correlation between mean metallicity and intrinsic metallicity spread among the objects, as seen in the right hand panels of Figure \ref{fig:distcomp}.  This correlation can be quantified by looking at the relative dispersion index (defined as $D = \sigma(Z)^{2}/\bar{Z}$) for each of the galaxies and star clusters.  This index can be used to infer the level of correlation between the variance and mean of a Z distribution; for a Poisson distribution, $D = 1$, while distributions with $D > 1$ and $D < 1$ are described by a negative binomial and binomial distribution, respectively.  
The dwarf galaxies and star clusters all show dispersion indices less than 1, and are therefore described by binomial distributions.  This observational result is interesting because these objects share the same functional form in their linear metal distributions, despite the range in masses and environments.

The globular clusters typically show the lowest values of $D$, ranging from $10^{-7} \leq D \leq 10^{-2}$, while the more luminous dwarf galaxies show distributions which are closer to a Poisson distribution, with the dispersion indices ranging from $10^{-3} \leq D \leq 10^{-1}$.  
Generally a binomial distribution can be thought of as a set of $n$ trials that occur with some probability of success $q$ (see $\S5.1$ below). This is a specific case of the more general Poisson-binomial distribution, which is a sequence of $n$ trials which each have an individual probability of success of $q_{i}$.  In the limit that all the probabilities are equal, then the Poisson-binomial reduces to the standard binomial distribution.  The interesting question of \emph{why} the distributions of linear metallicity all exhibit binomial forms will be discussed in $\S4$.

\section{Intrinsic Metallicity Spreads}
To analyze the intrinsic spread in metallicity for the objects in Table 1, we begin by removing the contribution to the metallicity spread due to observational errors.  To calculate the intrinsic metallicity spreads $\sigma([Fe/H])$ and $\sigma(Z)$, we follow the description by \cite{Kirby11a} to solve the following equation numerically for $\sigma([Fe/H])$:
\begin{equation}
\frac{1}{N}\sum_{i=1}^{N}\frac{([Fe/H]_{i} - \langle[Fe/H]\rangle)^{2}}{(\delta[Fe/H]_{i})^{2} + \sigma([Fe/H])^{2}} = 1.
\end{equation}
Where $[Fe/H]_{i}$ are the individual spectroscopic measurements from each star, and $\delta[Fe/H]_{i}$ are the individual measurement errors per star.  An analogous equation is used to solve for the intrinsic linear metallicity spread $\sigma(Z)$, with the the error in linear metal fraction computed from the [Fe/H] measurements and errors as:
\begin{equation}
\frac{{\delta}Z_{i}}{Z_{\odot}} = (Z_{i}/Z_{\odot})\rm{ln(10)}\delta[Fe/H]_{i}, 
\end{equation}


Errors on the intrinsic spreads are computed by numerical jackknife estimates, where each star is removed from the sample in turn, and the mean and variance recomputed.  The error on the mean or variance is then,
\begin{equation}
\sigma_{y} = \sqrt{\frac{N-1}{N}\Sigma(\bar{y}_{i} - \bar{y})^{2}},
\end{equation}
where $y$ is the quantity (mean, or variance) of interest.
In all cases, an upper limit to the intrinsic metallicity spread $\sigma$([Fe/H]) has been assigned to the distribution when the 
uncorrected metallicity spread has a smaller value than the smallest measurement error of an individual star, i.e., 
the upper limit value is taken as the smallest measurement error.  Table 2 shows the computed average metallicities and intrinsic metallicity spreads for the objects we consider.

\subsection{Consequences of the Binomial Nature of the Z Distributions for Interpreting Intrinsic Metallicity Spreads}
\begin{figure*}
\centering
\mbox{\subfigure{\ifpdf
\includegraphics[width=0.55\textwidth]{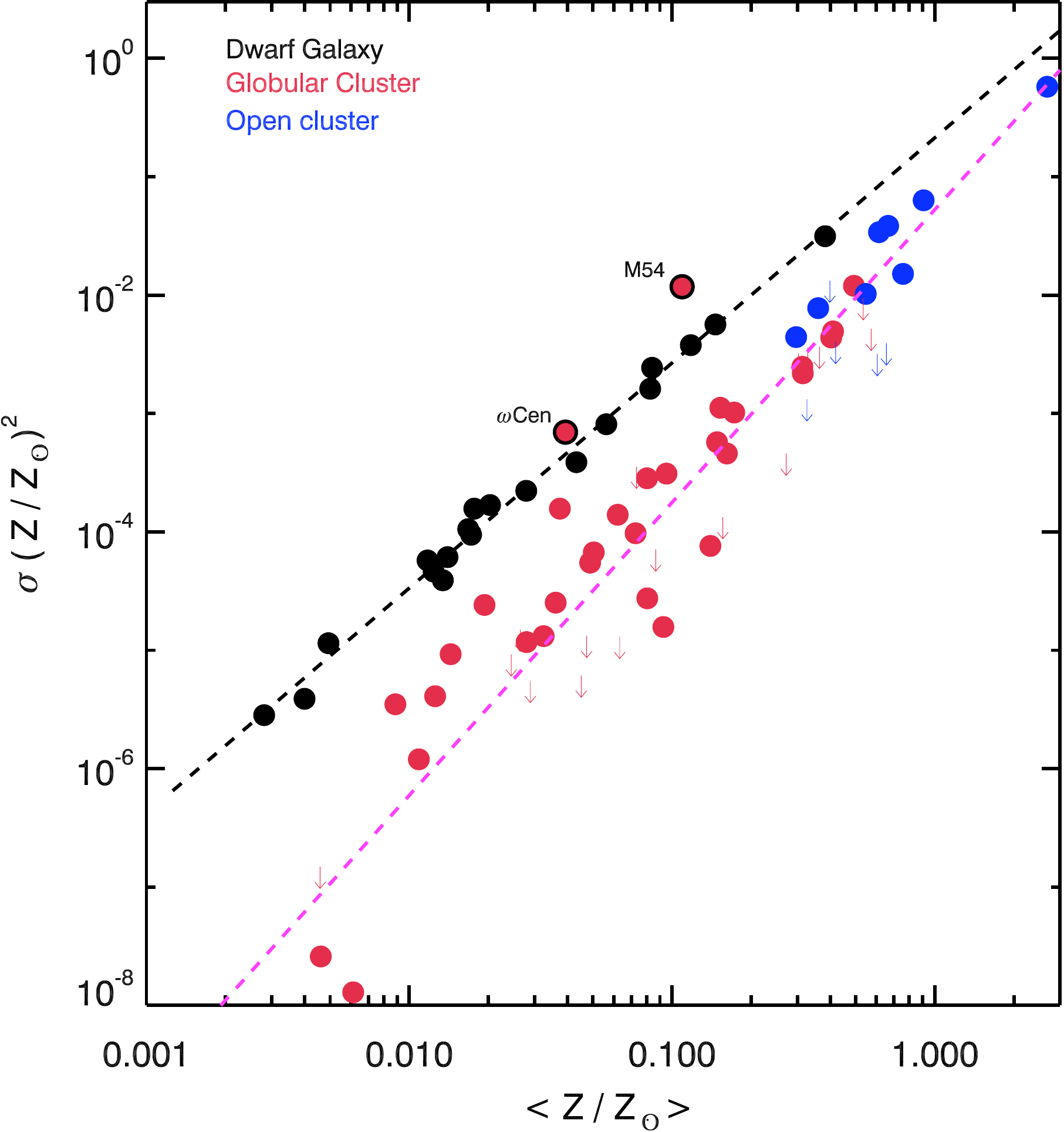}
\else
\includegraphics[width=0.58\textwidth,angle=-90]{meanvar2n.eps}
\fi
\quad
\subfigure{\ifpdf
\includegraphics[width=0.55\textwidth]{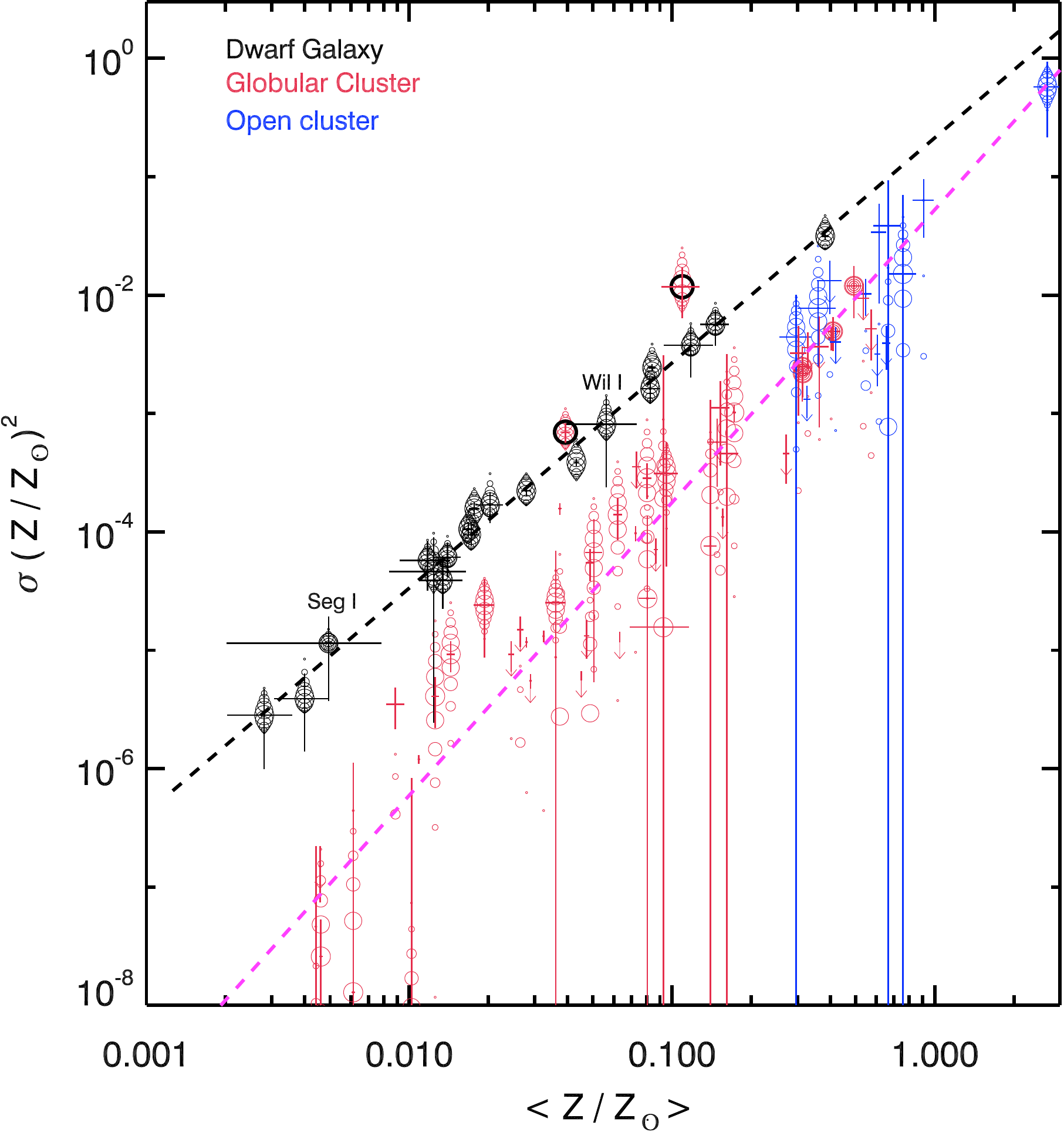}
\else
\includegraphics[width=0.58\textwidth,angle=-90]{meanvarerrn.eps}
\fi
 }}}
\caption{Intrinsic variance $\sigma(Z)^{2}$ in Z vs. $\bar{Z}$, the mean Z for dwarf galaxies and star clusters.  Dashed lines represent a linear least squares fit to the dwarf galaxies (\emph{black}) and star clusters (\emph{magenta}).  Arrows indicate upper limits to the intrinsic dispersions.   There is a clear separation between the dwarf galaxy and star cluster sequences. 
Right panel shows the same data but with jackknife sampling errors (solid lines) overlaid.  Open circles show the effects on the the intrinsic variance if the errors reported in the literature were under/over- estimated by $50 -150\%$ in $10\%$ increments for each object.}
\label{fig:meanvar}
\end{figure*}

To visualize the binomial nature of the linear Z distributions in the dwarf galaxies and clusters, Figure \ref{fig:meanvar} plots the intrinsic variance, $\sigma(Z)^{2}$, versus the mean linear metallicity, $\bar{Z}$.  There is a correlation between the mean metallicity and intrinsic spread in metallicity for both the dwarf galaxies and star clusters, characteristic of a binomial distribution as discussed in $\S3.1$.  
A clear separation in the sequences is evident, as dwarf galaxies at all metallicities show intrinsic metallicity spreads larger than the clusters at a given $\bar{Z}$.  The RMS scatter about the dwarf galaxy (DG) line is only 0.10 dex, while it is more than 7 times larger about the cluster sequence (0.71 dex).  Interestingly however, the star clusters do not scatter up and overlap with the galaxy population at any point.  This discrete offset is related to the fact that the star clusters have been pre-enriched within systems which themselves lie on the galaxy sequence.  Therefore the slope of the star cluster sequence, having the observational errors accounted for, represents a remnant signature of the self-enriched history which their formation environment (galaxies) underwent (see $\S5.3$).

The split in the DG and star cluster sequences is robust to both sampling errors and internal biases, i.e., 
if the quoted errors ($\delta[Fe/H]_{i}$) in the original studies are under or overestimated.
To illustrate this point, in the right panel of Figure \ref{fig:meanvar} the $\sigma(Z)^{2} - \bar{Z}$ diagram is shown with the error on each object from the jackknife tests.   
Objects with very few stars in them, such as Willman I and Segue I, show larger errors, but still remain offset from the 
star cluster sequence\footnote{We stress that this test is not sufficient to prove that Willman 1 and Segue 1 are dwarf galaxies 
since one or two Milky Way foreground stars in these small systems would enlarge their metallicity spreads so that they lie 
on the DG sequence.  
Proper foreground removal using statistical photometric and kinematic techniques is still crucial to understanding the nature
of these systems, e.g., \citealt{Aden09,Alan10,Koposov11,Munoz12}.}.  
The open circles in Figure \ref{fig:meanvar}, show the change in the intrinsic dispersion if the reported uncertainties 
on the [Fe/H] values, $\delta\rm{[Fe/H]}$, were under or over estimated by $50 - 150\%$ in $10\%$ increments.  
The sequences still show no overlap and there are no apparent correlations between the number of observed stars or quality 
of the observations.  The $\sigma(Z)^{2} - \bar{Z}$ appears to be a robust way to differentiate star clusters from dwarf galaxies, and therefore we suggest that this as a new and direct method for distinguishing these stellar populations.

\begin{figure}
\begin{center}
\ifpdf
  \includegraphics[width=0.99\textwidth]{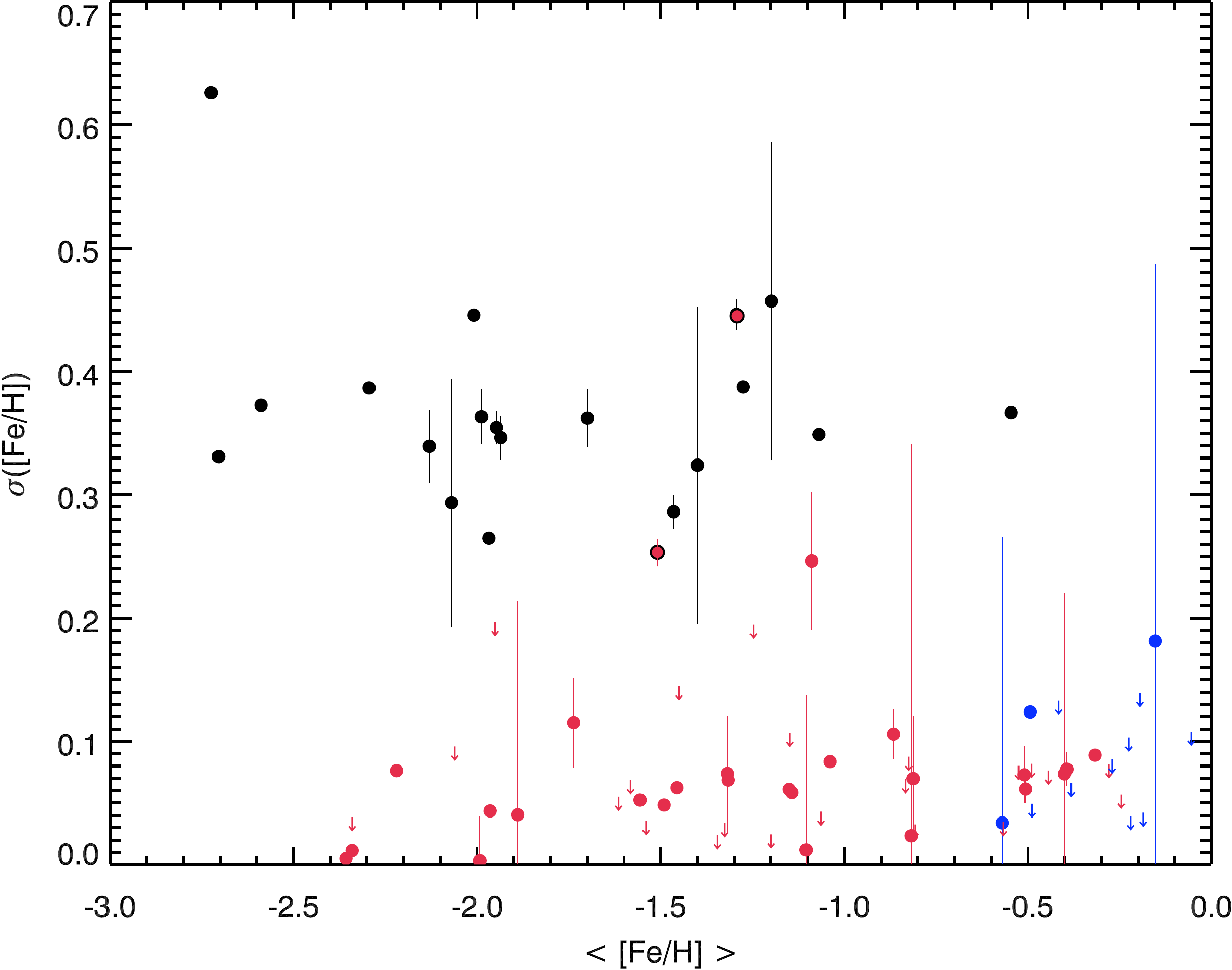}
  \else
  \includegraphics[width=0.85\textwidth,angle=-90]{meanvarfeh.eps}
  \fi
  \caption{Intrinsic spread in [Fe/H] versus average metallicity for dwarf galaxies and star clusters. This shows that there is no correlation between the mean [Fe/H] metallicity and the metallicity spread $\sigma$(Fe) 
in dwarf galaxies, unlike the correlation seen in Figure \ref{fig:meanvar}.
}
\label{fig:meanvarfeh}
\end{center}
\end{figure}

It is important to note that the correlation of intrinsic linear metallicity spread $\sigma(Z)^{2}$ with average metallicity $\bar{Z}$ is more nuanced than just assuming this is the natural result of how the metallicity increases with age of a system.  While systems with longer epochs of star formation will have larger spreads in metallicity, the DG sequence in Figure \ref{fig:meanvar} is \emph{not} a sequence in SFH (e.g., the positioning of Willman I is close to Fornax).  Similarly, if the expression of SFH were the dominant factor determining the correlations in Figure \ref{fig:meanvar} then one might expect such a correlation between metallicity spread and mean metallicity to hold in the [Fe/H] measurements.  However, as we show in Figure \ref{fig:meanvarfeh}, the intrinsic spread $\sigma([Fe/H])$ is independent of the average [Fe/H] of the dwarf galaxies.  
Thus the usefulness of the $\sigma(Z)^{2} - \bar{Z}$ diagram can be seen - 
this relationship is due to the discrete binomial form of the chemical evolution and linear metallicity distributions 
in these systems, and it is not simply a product of generic metal enrichment.

\subsection{Revisiting Metrics for Separating Star Clusters and Dwarf Galaxies}
\begin{figure}
\begin{center}
\hspace{2mm}
\ifpdf
\includegraphics[width=0.75\textwidth]{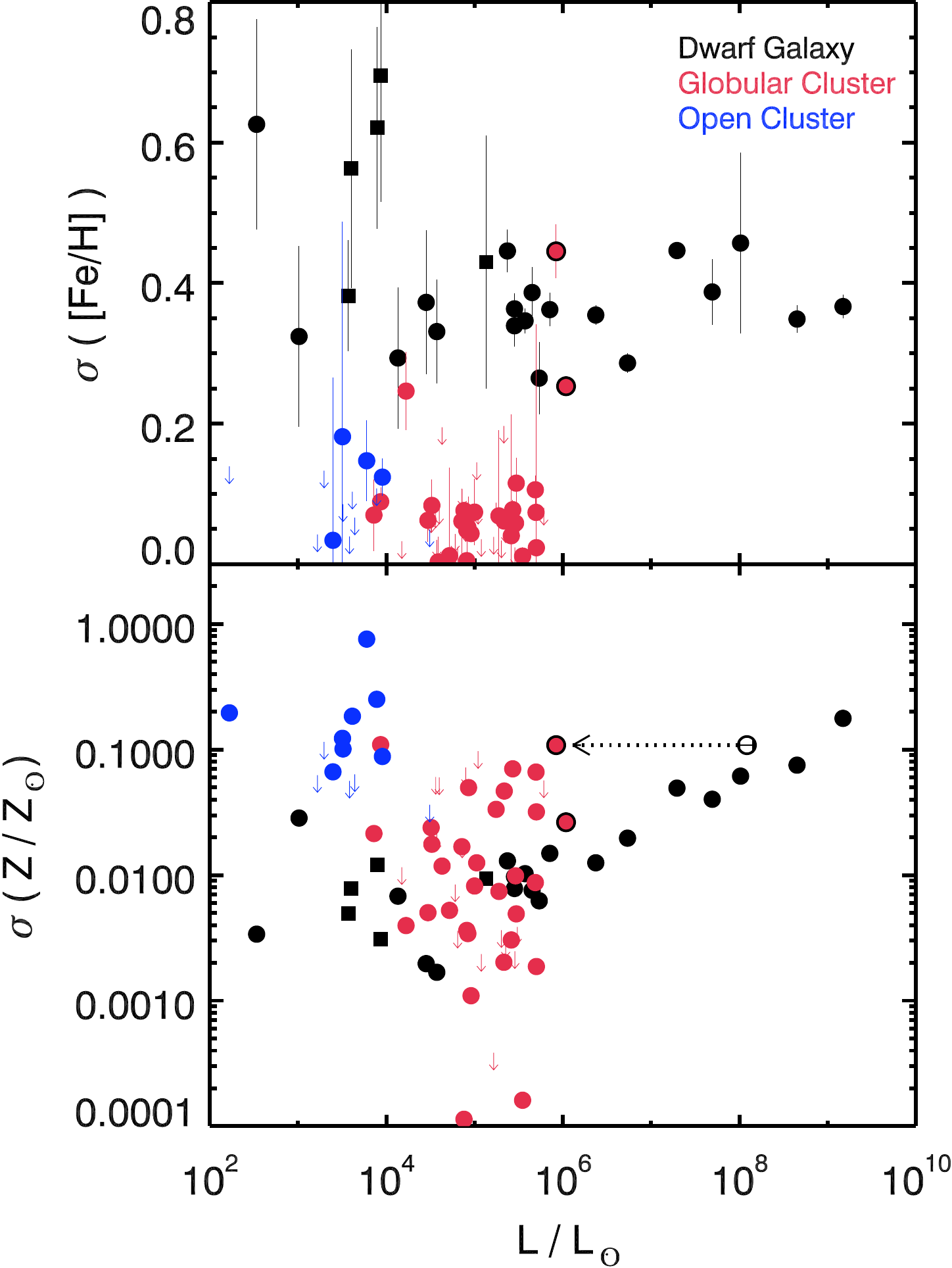}
\else
\includegraphics[width=0.99\textwidth,angle=-90]{sigzlumtest.eps}
\fi
\caption{Dispersion in [Fe/H] and Z versus host object luminosity.  Dwarf galaxies (\emph{black}) show a linear relation in $\sigma(Z)$ vs $L$ due to the linear metal fraction being binomially distributed and the well known luminosity-metallicity relation.  The 5 UFDs from the sample of \protect\cite{Kirby08} (black squares) have been added to show the linear anticorrelation exhibited below luminosities of $L = 10^{5} L_{\odot}$.  Red points with black circles are M54/Sgr and $\omega$Cen, which are both likely associated with former dwarf galaxies.  The open black circle represents recent estimates of the initial luminosity of the Sgr dSph from the study of \protect\cite{Niederost12}.  Small arrows indicate upper limits.}
\label{fig:siglum}
\end{center}
\end{figure}

With the intrinsic value of the metallicity spread, $\sigma([Fe/H])$ or $\sigma(Z)$ and the associated errors, we compare the behaviour of the metallicity dispersions of the star clusters, and dwarf galaxies in the $\sigma([Fe/H])-L$ and $\sigma(Z)-L$ planes in Figure \ref{fig:siglum}.  As expected, the clusters show very low spreads in metallicity relative to the dwarf galaxies (c.f., \citealt{CG09,Willman12}).  
However in the top panel of Figure \ref{fig:siglum} the high luminosity systems show a flat trend in $\sigma([Fe/H])-L$, and as expected the M54/Sgr\footnote{We consider all stars in the M54 and Sgr region from this study and note that due to the extreme difficulty in associating stars simply to M54 or Sgr (see \citealt{Carretta10}), the values in this table should be considered upper limits} and $\omega$Cen are the only  potential clusters that show a large spread in heavy metals - consistent with the theory that  both may be the nuclear remnants of accreted dwarf galaxies (e.g., \citealt{Bekki03,Carretta10}).

In the bottom panel of Figure \ref{fig:siglum} the dwarf galaxies show a tight correlation in $\sigma(Z)-L$, as first reported by \cite{Kirby11a}.  Interestingly, the globular clusters and open clusters do not show this trend, and yet the former overlap in the parameter space of the dwarf galaxies with luminosities of $10^{5} - 10^{6} L_{\odot}$.   
The linear metallicity spread of an object in this luminosity range is then not necessarily indicative of it being a galaxy.

Given the well known luminosity-metallicty relation exhibited by Local Group dwarfs \citep{DekelWoo03,Kirby11a}, plotting $\sigma(Z)$ versus $L$ is equivalent to $\sigma(Z)$ versus average metallicity $\bar{Z}$.  As discussed in $\S 3$ the Z distributions are clearly non-Gaussian, and in fact they are binomial.
Therefore it is not surprising that a tight sequence is recovered, as the variance and mean are proportional in a binomial distribution - and the $\sigma(Z)-L$ plot is simply recovering the autocorrelation between the mean and the variance of the dwarf galaxy $Z$ distributions.  This tight sequence in the dwarf galaxies is contrasted by the behaviour of the star clusters, 
which scatter broadly over a range of metallicity spreads.  The star clusters do not show a correlation in $\sigma(Z)-L$ since there is no metallicity-luminosity relation for GCs in the Milky Way (c.f., \citealt{BH09}).  

Clearly $\sigma(Z) - L$ is not an ideal metric for differentiating star clusters and DGs, and 
Figure \ref{fig:meanvar} offers a robust alternative in $\sigma(Z)^{2}-\bar{Z}$.    This metric does not require 
knowledge of the host object's luminosity, or mass, but simply a distribution of metallicities.  
While useful in classifying low luminosity systems, the binomial nature of the metallicity distributions and the $\bar{Z} - \sigma(Z)^{2}$ correlations may also allow for information on chemical enrichment processes to be obtained when compared to suitable models.     

\section{Discussion}


We have demonstrated that due to the binomial nature of the linear metallicity distributions for star clusters and dwarf galaxies, 
separating them in the $\sigma(Z) - L$ plane is difficult since each class of object has a different luminosity-metallicity 
relationship.  A much cleaner distinction between star clusters and DGs is found when plotting the intrinsic spread 
in linear metallicity (characterized by the intrinsic variance) as a function of its mean metallicity, 
$\sigma(Z)^{2} - \bar{Z}$.   In this parameter space, dwarf galaxies form a tight sequence over nearly 7 orders of 
magnitude in luminosity, and separate out cleanly from both open clusters and globular clusters.  
Furthermore, the correlation between $\bar{Z} - \sigma(Z)^{2}$ is natural -  a physical result of the sequential 
buildup to present day metallicities through \emph{discrete} enrichment events in a galaxy that are small in magnitude 
relative to the total metallicity evolution.    Understanding the relative scatter and offset in these two sequences 
offers new information on the different modes of chemical enrichment expected to dominate the two classes of objects.

\subsection{Linking Binomial Parameters to Physical Processes in SF}
Having shown that the linear metal distributions are described by binomial distributions, it is important to associate the binomial distribution parameters to physical processes in the chemical enrichment of a galaxy.  As discussed in $\S3.1$, the number of successes $n$ and probability of each success $q$ which characterize a binomial distribution can be thought of as the number of star formation or enrichment ``successes'' in a galaxy or star cluster, which occur with a characteristic probability.  Given the clean separation when looking at the dispersion indices, it would be useful to apply analytic chemical evolution models to the data that incorporate the binomial form of the $Z$ distributions, while providing a link between the statistical properties of the distributions and physical processes related to enrichment.  There are literature studies that have described galaxy chemical evolution using stochastic and inhomogeneous chemical enrichment models (e.g., \citealt{vandenhoek97,Argast00,Karlsson05,Salvadori09}) 
that can be compared to discrete statistical distributions. 
In particular, \cite{Oey00} described a modified simple one-zone chemical evolution model based on binomial parameters which physically corresponded to the number of enrichment generations, $n$, and the filling factor of chemical enrichment within the galactic ISM, $q$. In that picture the binomial ``successes'' are physically realized as the successive enrichment of a particular volume of gas, which provides an excellent physical link between chemical evolutionary processes and the binomial Z distributions found here.

\subsubsection{\protect\cite{Oey00} Inhomogeneous Chemical Evolution Model}
In the simple closed system that \cite{Oey00} considered, the inhomogeneous enrichment of a galaxy or star cluster proceeds through sequential pollution of regions produced by SNe.  For $n$ generations of star formation and subsequent enrichment, the galaxy is polluted by bubbles that have some covering fraction $q$ of the whole galaxy.  Within a given bubble the yield of metals from SNe is assumed to be constant, and therefore the metallicity within a given volume is determined only by the size that the bubble sweeps out - as the region's metallicity is the result of diluting the pristine ISM with the SN ejecta.  

For a single generation of star formation, the distribution of bubble sizes (and thus metallicity of the regions) is well described by a power law, $N_{s} \propto R^{1-2\beta}$ with slope $\beta = 2$ \citep{Oey97}.  
The distribution of metals then follows from this distribution of bubble sizes:
\begin{equation}
f(z) \propto \frac{4}{3}{\pi}R^{3}N_{s}(z) \propto z^{-2}dz;~~~~~~~z_{min} \leq z \leq z_{max},
\end{equation}
up to some minimum and maximum metallicity.  These limits on the $Z$ value of the regions in a given generation can alternatively be thought of as corresponding limits on the region size, $r_{min}$ and $r_{max}$.  \cite{Oey00} bounded the minimum bubble size to be that of observed supernovae remnants, $\sim 25$pc in the Milky Way.   We relax this constraint in $\S5.2$ to fully describe the binomial self enrichment in terms of the physical properties of a galaxy's ISM.

\begin{figure}
\begin{center}
\ifpdf
\includegraphics[width=0.99\textwidth]{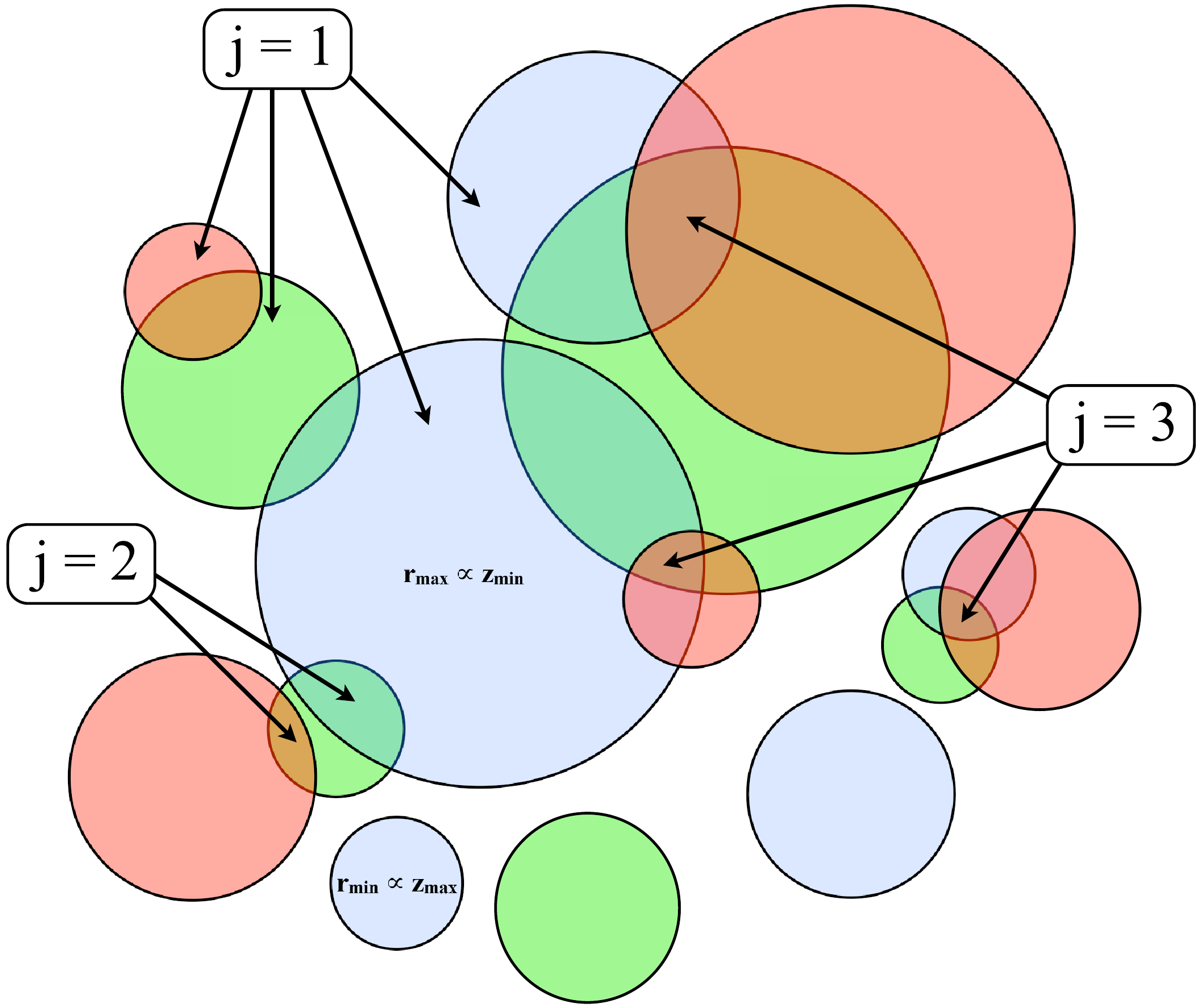}
\else
\includegraphics[width=0.85\textwidth,angle=90]{oeymod.eps}
\fi
\caption{Illustration of the \protect\cite{Oey00} binomial chemical evolution model (adapted from Fig. 1 in \protect\cite{Oey03}).  Three generations of enrichment ($n=3$; green, red, and blue) are shown, each have a covering fraction of $q$.  The range in bubble sizes, from $r_{min}$ to $r_{max}$ correspond to a range in metallicities ($Z_{min},Z_{max}$) within each bubble assuming the constant yield of metals within a region is diluted in the corresponding volume.  The probability that any point is enriched by $j$ overlapping regions is given by the binomial probability $P = {n \choose j} q^{j}(1-q)^{n-j}$.}
\label{fig:oeymod}
\end{center}
\end{figure}

At any given location in the galaxy, the instantaneous metallicity depends on the number of SF generations $n$ and the covering fraction $q$ of the polluting bubbles.  This location may be polluted by one metallicity region, or it may be overlapped by a second generation, or overlapped by up to $j$ regions.  \cite{Oey00} adopted the likelihood that any location is overlapped by $j$ regions is given by the binomial probability:
\begin{equation}
 P_{j} = {n \choose j}q^{j}(1-q)^{n-j};~~~~~~ 1 \leq j \leq n.
\end{equation}
In Figure \ref{fig:oeymod} we show a schematic representation of how the number of generations, amount of overlap, and range of bubble sizes can be visualized in this framework (taken after \citealt{Oey03}).

The total metallicity distribution function of a host system after $n$ generations is then:
\begin{equation}
N_{tot}(z) = \frac{1}{n}\sum^{n}_{j=1}\sum^{n}_{k=j}D_{k-1}P_{j,k}N_{j}(z),
\end{equation}
Where $D$ is a factor to account for the consumption of gas in the system (see Section 2 of \citealt{Oey00} for more details).
With these equations from the \cite{Oey00} model,
it is possible to create simulated populations for a given $n$ and $q$ using Monte Carlo methods, 
which has the binomial nature of the Z distributions encoded and directly linked to physical processes in chemical evolution.

\subsection{Application of the \protect\cite{Oey00} Model to the $\sigma(Z)^{2} - \bar{Z}$ Diagram}
The \cite{Oey00} binomial chemical evolution model describes the physical processes that control the placement 
and evolution of star clusters and dwarf galaxies in the $\sigma(Z)^{2} - \bar{Z}$ plane.   The specific 
definition of the model parameters could be refined in more complex chemical evolution models (e.g., low $q$ could be analogous to describing the fraction of ejected/outflowing metals), however here we consider only a general model for our simple applications.  In addition to the number of SF generations, $n$ and covering fraction $q$, the final metallicity distribution in the model depends on the lowest metallicity of any region, $Z_{min}$, and the range of metallicities ${\Delta}Z \equiv Z_{max}-Z_{min}$.  Thus ${\Delta}Z$ is analogous to describing the range of characteristic region sizes, $r_{min}$ to $r_{max}$ found in a galaxy, such that small ${\Delta}Z$ would require a system to have a limited range of bubble sizes (which may be the case when the object is intrinsically small as in the case of a star cluster).


With a choice of initial limits for these bubble sizes/metallicities, the prescriptions in the \cite{Oey00} equations permit one to take an object from a starting point on the $\sigma(Z)^{2}-\bar{Z}$ diagram, and chemically evolve it an arbitrary amount.  The resultant time-integrated MDF of the object can be computed through Monte Carlo simulations of mock populations using Equations 4-6. 
We have computed a grid of simulated MDFs that chemically evolve a host from various starting points to quantify the behaviour of objects within this model framework.  At each end point of a simulated system, the final time-integrated MDF was output, and the intrinsic variance and mean of the MDF measured in order to compare to the observational data.  The 4-dimensional grid of models has a variance and mean of the time integrated MDF computed at every point, with the grid points as follows:  $n = [1,2,5,10,25,55]$ generations; covering fractions of $q = [0.1,0.25,0.5,0.75,0.9,1.0]$; $Z_{min} = [10^{-3.7},10^{-3.0},10^{-2.3}]$, and ${\Delta}Z = [0.3,0.8,1.6]$ dex.  We present subsets of the model tracks in Figure \ref{fig:meanvarmodt} with two parameters fixed, and two varying in each of the panels.

As noted by \cite{Oey00}, the model is partially degenerate between $n$ and $q$ such that the product of those two parameters describes the degree to which a system is chemically evolved.  This degeneracy can be seen in the top two panels of Figure \ref{fig:meanvarmodt} when comparing the metallicity spreads and means.  In both cases as the product $nq$ increases, the final object moves to a region of higher mean metallicity and metallicity spread, with the binomial model doing an excellent job of naturally reproducing the slope of the DG sequence.  The lower left panel shows that the effect of changing $Z_{min}$ predictably shifts any starting point along the x-axis.   
The lower right panel illustrates that increasing ${\Delta}Z$ moves the start point to higher metallicity spreads.  
Notice that for the DGs to have started on their sequence, they require a larger range of metallicities within their 
enrichment regions (which may be determined by low mixing efficiencies), or equivalently would need a range of bubble sizes larger than what is needed for starting on the GC sequence.
This is consistent with simple expectation for dwarf galaxies, where intrinsic sizes are larger 
($>25$ pc; \citealt{Gilmore07,Tolstoy09, Alan12}) 
than the minimum bubble size corresponding to the smallest region adopted by \cite{Oey00} ($<$25 pc).

The variation in starting points as a function of ${\Delta}Z$ is an important point to consider, as it explains how dIrrs with $\sim 12$ Gyr of continual SF (such as the LMC), can lie on the same sequence as dSphs like Segue 1 or Tucana which had all of their stars form in $\leq 1$ Gyr \citep{Tolstoy09}.  If the number of enrichment events (or equivalently SF duration) was the only parameter determining a galaxy's positioning on Figure \ref{fig:meanvarmodt} it would be difficult to reconcile how such diverse galaxies could lie on a similar sequence.  However, the lower right panel illustrates that for specific ranges of ${\Delta}Z$ a galaxy could start and end on the present day DG sequence even with a brief SFH.  It should be kept in mind that the application of the \cite{Oey00} model is most useful in a differential sense - i.e., the absolute value of ${\Delta}Z$, and therefore the physical range of initial DG sizes, will depend somewhat on the yield assumed from SNe.
   
The model tracks in Figure \ref{fig:meanvarmodt} indicate that objects may never move horizontally - and are restricted to never evolve flatter than the slope of the current DG sequence.  
The restricted changes in the metallicity spread of a system places a fundamental limit on the homogeneity of the chemical enrichment in dwarf galaxies.  The predictions from the model match the observed DGs, and are consistent with the canonical theory of self-enrichment dominating in dwarf galaxies. The fact that this holds over 8 orders of magnitude in dynamical mass suggests that while the enrichment details differ, the dwarf galaxies themselves undergo similar inhomogeneous chemical evolution (or merging; c.f. \citealt{Salvadori09}) processes throughout their lifetime - with their end states well described by a simple binomial model. 

\subsection{Star Cluster Pre-Enrichment and Dwarf Galaxy Self-Enrichment in the $\bar{Z} - \sigma(Z)^{2}$ Diagram}
If the chemical evolution in the dwarf galaxies was dominated by self-enrichment, whereas the star clusters are pre-enriched, then it is possible to describe both the split in the sequences in Figure \ref{fig:meanvar} and the larger scatter in the star cluster sequence in the framework of the inhomogeneous mixing model.  In this picture the proto-GCs would have had their ISM enriched by the larger environment (a galaxy host) out of which they formed, with subsequent SF being too brief, too energetic, or the gas content simply too small to allow for continued enrichment within the cluster.  The smaller values for $\sigma(Z)^{2}$ at all metallicities for the star clusters would be expected then, as they would form stochastically from portions of a galaxy's ISM that has been inhomogeneously enriched.  On the other hand, dwarf galaxies would start their chemical evolution from a relatively un-enriched ISM and can undergo a larger change in mean metallicity (increasing $\sigma(Z)^{2}$) due to more efficient retention of metals and gas in their larger potential wells.  An increase in $\sigma(Z)^{2}$ for the dwarf galaxies would also be produced if they are fed with un-enriched gas via cold flows \citep{Keres05,Brooks09}.  It is important to note that the duration of star formation is likely not the driving parameter between the two sequences as the SF epoch in the clusters is thought to be nearly instantaneous, but the SFHs among the dwarf galaxies range from less than a Gyr to a Hubble time and yet they still lie on the same sequence.  This is most apparent in considering the location of Willman I, one of the least luminous systems, which occupies a region very near to the SMC and Fornax.

\begin{figure*}
\begin{center}
\ifpdf
\includegraphics[width=0.99\textwidth]{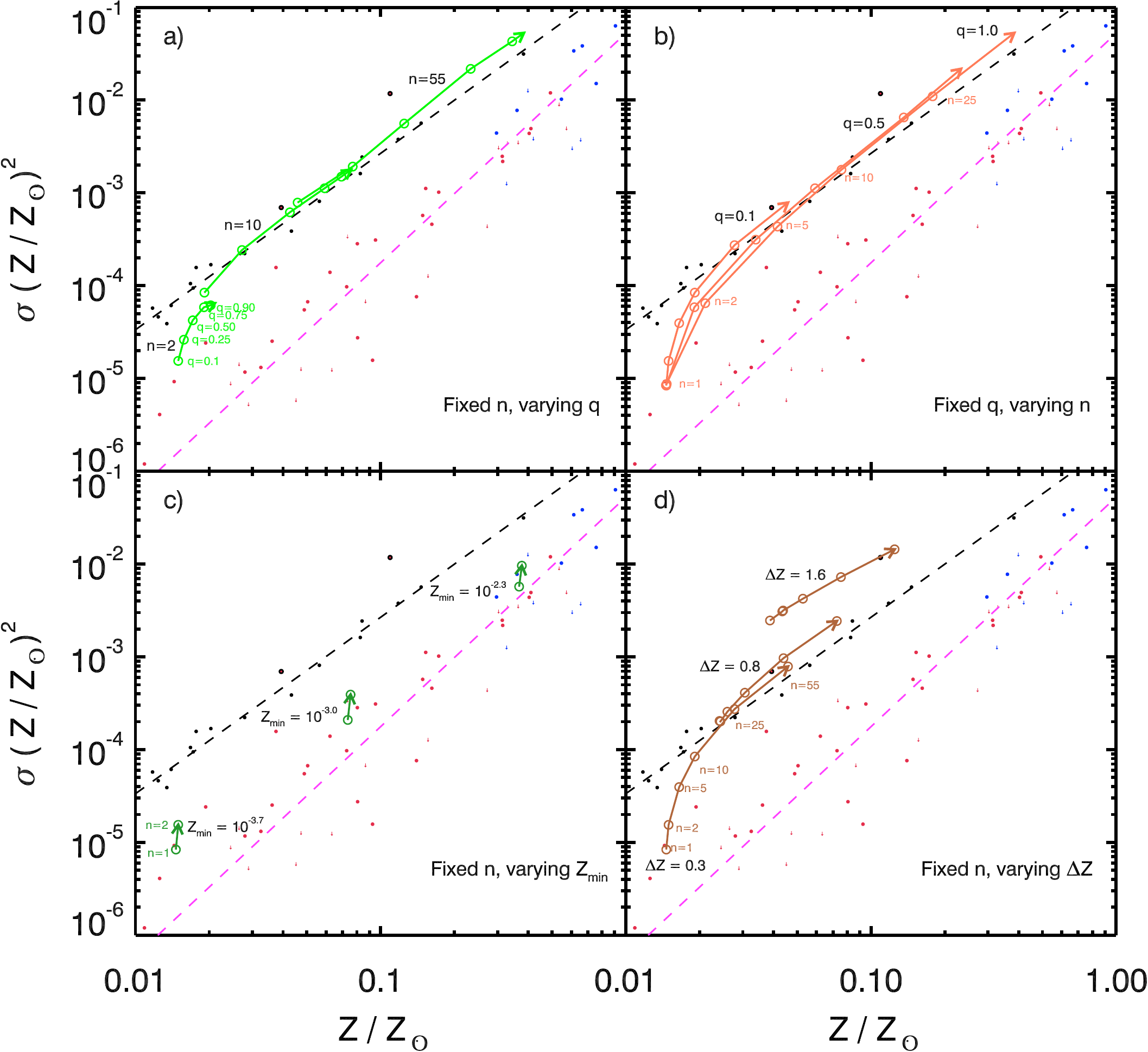}
\else
\includegraphics[width=0.90\textwidth,angle=-90]{meanvarmodt.eps}
\fi
\caption{Change in the variance and mean for different combinations of $n$ number of generations, $q$ covering fraction, minimum bubble metallicity $Z_{min}$, and range of region metallicities ${\Delta}Z$ taken from the binomial chemical evolution model of \protect\cite{Oey00}.  Lines show the computed variance and mean of simulated linear metallicity distributions as the models are evolved from $n=1$ to $n=55$.  Black, red and blue points show the same dwarf galaxy, globular cluster, and open cluster data from Figure \ref{fig:meanvar}.  The lower panels show that the clusters must have been pre-enriched, unlike the dwarf galaxies which demonstrate strong self-enrichment in the upper panels.}
\label{fig:meanvarmodt}
\end{center}
\end{figure*}

The slope of the DG $\sigma(Z)^{2}-\bar{Z}$ sequence is naturally reproduced as the product of $nq$ increases in the inhomogeneous model of \cite{Oey00} (as discussed in $\S5.2$).  
The limiting slope produced by the models requires either that DGs have been born on the sequence 
with a larger ${\Delta}Z$ value, or that they start below it but self enrich significantly 
(panel d, Figure \ref{fig:meanvarmodt}).  
This can set complementary constraints on the initial sizes of DGs early in their history, or on the number of stellar generations.  If all DGs started at a value of ${\Delta}Z = 0.8$ as shown in panel d of Figure \ref{fig:meanvarmodt}, this may be informing us about the characteristic clustering size of the central regions of their DM halos.  
As discussed by \cite{Gilmore07}, if the baryons in DGs are initially distributed such that their characteristic scale length is set by the minimum dark matter clustering length or core size, then this may set $q$ and ${\Delta}Z$.  Alternatively the mid-plane ISM pressure has been suggested to have a dependence on SFR \citep{Shetty12}, which may result in mixing efficiencies which are common to most galaxies.  In both scenarios chemical evolution proceeds in a singular direction due to similar initial conditions or ISM physics, and results in the slope and narrow scatter seen in Figure \ref{fig:meanvar}.  Thus, the small scatter suggests that dwarf galaxies have been dominated by self enrichment, consistent with the expectations for their chemical evolution.

By contrast, self-enrichment among the star clusters would be inconsistent with the large scatter in their $\sigma(Z)^{2} - \bar{Z}$ sequence.  Panel c of Figure \ref{fig:meanvarmodt} illustrates the effect of a second generation of SF on the GCs.  The increased spread in metallicity is far less than the scatter shown by the cluster points, suggesting that even with a second generation self-enrichment cannot fully explain the star clusters sequence.  
In addition the direction of movement after one generation at such chemically unevolved states is nearly vertically.  
This importance of ${\Delta}Z$ can also be appreciated when asking if GCs can move diagonally along their sequence.  
The only way to start an object on the GC sequence is to change ${\Delta}Z$ to low values.  However this not only starts the evolutionary model at a lower metallicity spread, but simultaneously restricts the subsequent generations to increase in dispersion faster than mean metallicity.

\begin{figure}
\begin{center}
\ifpdf
  \includegraphics[width=0.99\textwidth]{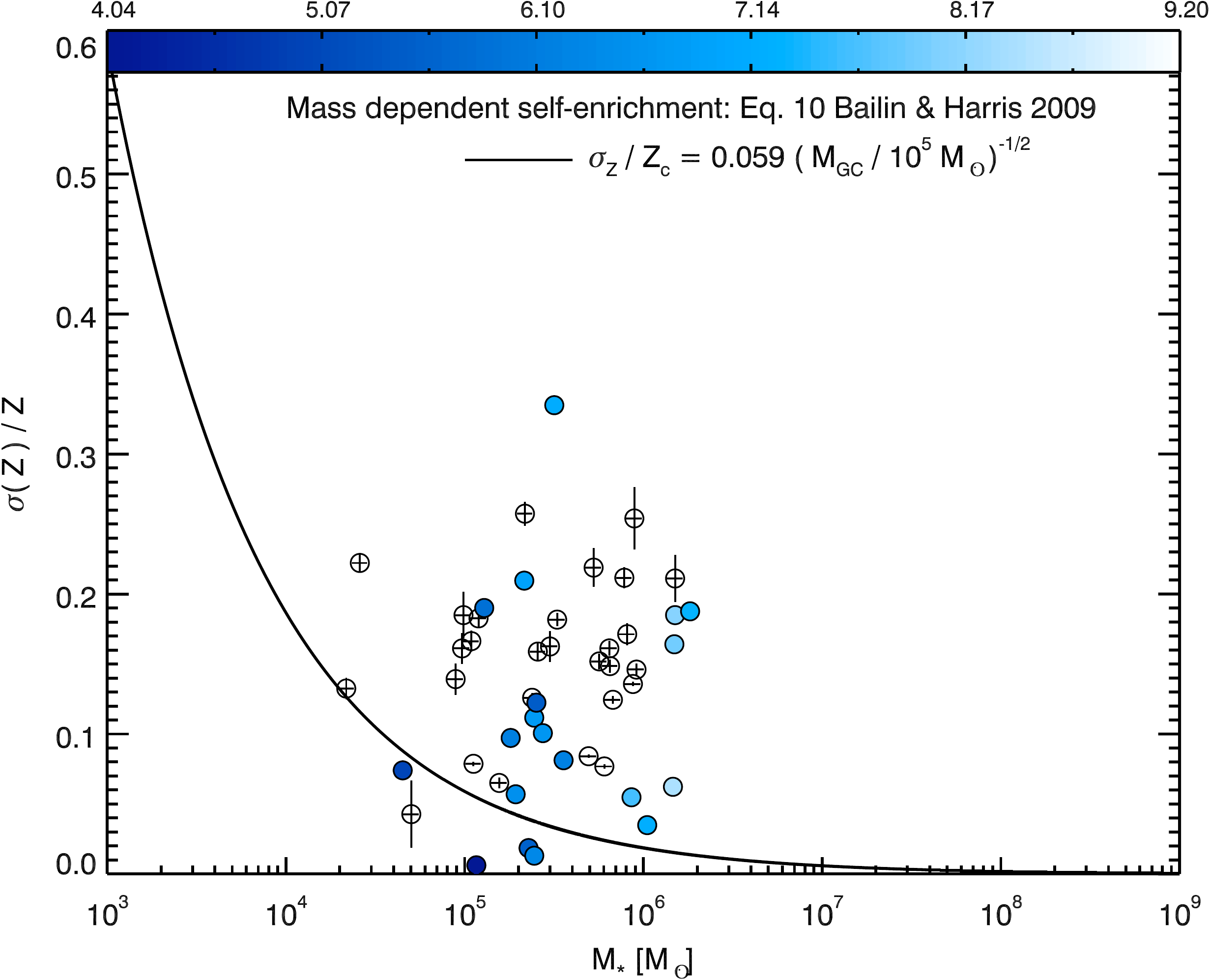}
  \else
  \includegraphics[width=0.85\textwidth,angle=-90]{mgcdis.eps}
  \fi
  \caption{The relative spread in Z as a function of globular cluster mass.  The theoretical upper limit to globular cluster self enrichment as a function of proto-cluster cloud mass (from \protect\cite{BH09}) is shown as the solid line.  Shading corresponds to total cluster mass enhancement at formation relative to the present time, as estimated from the Na-O anticorrelations by \protect\cite{Conroy11}.  Incorporating the mass enhancement multipliers, all clusters but NGC 6739 (darkest blue circle) show metal spreads larger than can be accounted for by self-enrichment.}
\label{fig:mgcdis}
\end{center}
\end{figure}

A further elegant constraint on the amount of self-enrichment in GCs was demonstrated by \cite{BH09}, who  
showed that the relative spread in cluster metallicity could be expressed independent of star formation or metal retention efficiency and only as a function of cluster mass:
\begin{equation}
\frac{\sigma_{Z}}{Z_{c}} = 0.059\left( \frac{M_{GC}}{10^{5} M_{\odot}}\right)^{-1/2}.
\end{equation}
This maximum amount of self-enrichment ($\sigma(Z)/Z_{c}$) that a globular cluster of a given mass could undergo provides a strict upper limit to the cluster's ability to enrich, and can be easily compared to the observed ratio of $\sigma(Z)/\bar{Z}$.  In Figure \ref{fig:mgcdis} we show the relative metallicity spread $\sigma(Z)/\bar{Z}$ for the globular cluster points as a function of cluster mass, where masses have been computed assuming the luminosity is converted with a constant mass-to-light ratio of 2 (i.e., \citealt{Alan12}).  The solid curve shows the maximum metallicity evolution ($\sigma(Z)/Z$) produced via self-enrichment, as a function of star cluster mass.
Except for five of the clusters, all GCs in the sample  show relative metallicity spreads that are too large to be produced by self enrichment.  This provides an independent check on the ability of GCs to self enrich, confirming the results from the application of the \cite{Oey00} model to the offset sequences in Figure \ref{fig:meanvarmodt}.  

The inability of self-enrichment to dictate a star cluster's evolution is even more striking when one accounts for the fact that most if not all clusters have had two generations of SF\footnote{A second generation is required to explain the elemental anticorrelations and morphological features in the colour magnitude diagrams of MW GCs, where the second generation stars are polluted by the ejecta from AGB stars formed in the first generation (c.f., \citealt{Gratton12}).}.  \cite{Conroy11} showed that ejecta from the primordial generation of stars could only be produced and retained if the globular clusters were many times more massive than their present day stellar mass.  Using a diffusive model, \cite{Conroy11} computed the mass enhancement factor ($f_{t} = N{\times}M_{present}$) for the clusters based solely on the strength of their observed Na-O anticorrelations.  We have colour coded the globular clusters in Figure \ref{fig:mgcdis} with the mass enhancement values from \cite{Conroy11}, and it is clear that four of the five clusters in Figure \ref{fig:mgcdis} which currently lie below the solid self-enrichment limit would have had original masses that place them above that curve.  This further suggests that globular clusters are inconsistent with having been strongly self-enriched.

If globular clusters are unable to efficiently self-enrich to explain their $\sigma(Z)/Z$ values, the alternative is that their positioning and scatter on the $\sigma(Z)^{2}-\bar{Z}$ plane is set by pre-enrichment.  This requires a galaxy to self-enrich inhomogeneously, such that the location or time of formation of a GC from that galaxy sets the GC mean metallicity and intrinsic spread.  In this picture of pre-enriched cluster formation, a parcel of gas within a self-enriched galaxy could form a GC at any point during the galaxy's chemical evolution - however due to the inhomogeneous mixing implicit in the binomial model, the stochastic sampling of the host galaxy's ISM would lead to smaller $\sigma(Z)^{2}$ values for the GC than for the host DG.  This process results in the offset of the star cluster sequence in the $\sigma(Z)^{2}-\bar{Z}$ diagram.  The slope of the cluster sequence is then a remnant signature of the self-enriched history of the galaxy's gas parcel out of which the GC formed, as the clusters are unable to evolve diagonally as shown in panel d of Figure \ref{fig:meanvarmodt}. This picture of GC pre-enrichment is also consistent with the lack of strong correlation between deviations from the best fit cluster sequence in Figure \ref{fig:meanvar} and current host properties of GCs - as their pre-enrichment levels were dictated by the host galaxy.

\subsection{Future Uses for the $\sigma(Z)^{2}-\bar{Z}$ Relationship}
Analysis of the star cluster and dwarf galaxies jointly in the $\sigma(Z)^{2}-\bar{Z}$ diagram in the context of binomial self-enrichent and pre-enrichment may be used to constrain the early evolutionary history of the dwarf galaxies and star clusters, but also to study how the system of Milky Way star clusters were accreted and/or formed in situ.  
Under the assumption that all GCs originated in galaxies that were self-enriched, then the \cite{Oey00} model in conjunction with $\sigma(Z)^{2}-\bar{Z}$ values place constraints on the possible physical sizes or chemically evolved states of the progenitor galaxies.  This ability to track the stochastic buildup of metallicity over time with the \cite{Oey00} model, allows one to quantitatively describe the formation environment for star clusters - something not possible with usual analytic chemical evolution models.  This could be powerful when used with studies that place simulated dwarf galaxies on the $\sigma(Z)^{2}-\bar{Z}$ plane.  Fine tuning of feedback or gas inflow/outflow efficiencies in those simulations to reproduce the dwarf galaxy sequence could result in further insights into the physical processes which dictate the slope and scatter in that parameter space. 

\section{Conclusions}
We have analyzed the metallicity distributions of a sample of open and globular clusters together with dwarf galaxies in the Local Group, focusing on the statistical nature of their linear metallicity ($Z$) distributions.  Below is a brief summary of the main findings from this analysis:
\begin{itemize}
\item The star clusters and dwarf galaxies show strong correlations between their mean linear metallicity ($\bar{Z}$) and intrinsic spread in linear metallicity ($\sigma(Z)^{2}$).  These correlations are consistent with a binomial distribution, (i.e., where $\sigma(Z)^{2}/\bar{Z} < 1$), unlike $\sigma([Fe/H])^{2}$ which has no correlation with $\bar{[Fe/H]}$
\item Plotting $\sigma(Z)^{2}$ against $\bar{Z}$ for all the objects shows two distinct sequences - one for the star clusters and another for the dwarf galaxies.   The $\sigma(Z)^{2} - \bar{Z}$ parameter space is therefore useful at distinguishing 
dwarf galaxies from star clusters, without additional information (such as $L, M_{dyn}$).  We find that $\omega$Cen and M54/Sgr 
fall on the dwarf galaxy sequence, confirming expectations that they represent the nuclei of accreted dwarf galaxies.
\item By applying a binomial chemical evolution model to objects in the $\sigma(Z)^{2} - \bar{Z}$ parameter space (e.g., Oey 2000),
the slope and offsets of the dwarf galaxy and star cluster sequences can be reproduced and explained in terms of the number of 
star formation generations, the covering fraction, and characteristic size of the enrichment events.
\item Within this model framework, we illustrate how the inhomogeneity of the self-enrichment process for galaxies dictates the slope of their sequence in the $\sigma(Z)^{2}-\bar{Z}$ plane.  The offset of the star cluster sequence reflects their formation by stochastic pre-enrichment from a parcel of gas inhomogeneously mixed within their host galaxy.  The star cluster slope  represents a remnant signature of the self-enriched chemical history of the galaxies from which they formed.
\item This model confirms expectations that star clusters cannot have been strongly self-enriched like the dwarf galaxies, and that pre-enrichment must have been the dominant mechanism affecting the star clusters' heavy element properties.  
This model also explicitly describes the connection between the two enrichment modes of galaxies and star clusters, in a self consistent and physically motivated framework.
\end{itemize}
Future applications of the \cite{Oey00} model may be able to provide constraints (size, ISM density, potential well depth) on the physical environment of the high redshift galaxy environments that star clusters are sampling.   Also, while the model is simplistic, constraints on enrichment and SF physics may be studied by comparing the predictions from high resolution SPH galaxy simulations in the $\bar{Z} - \sigma(Z)^{2}$ parameter space to observations.   Through such comparisons, variations produced by changing SF or feedback prescriptions in the simulations could be directly related to movement in $\bar{Z} - \sigma(Z)^2$ diagram.
\acknowledgments
We would like to acknowledge comments from the anonymous referee which greatly improved this manuscript.  The author would like to thank Kim Venn for careful reading of this manuscript, J. Trevor Mendel, Alan McConnachie, Sally Oey, Else Starkenburg, Giuseppina Battaglia and Alyson Brooks for useful discussions, as well as the members of the UVic Stellar Astrophysics group for helpful suggestions in preparation of this work.  The author acknowledges support from NSERC Discovery Grants to Kim Venn. The author would like to thank all authors for making their data public, and Ivo Saviane for providing additional data prior to publication.

\bibliography{varz}
\clearpage
\clearpage
\clearpage
\clearpage
\clearpage
\clearpage
\clearpage

\clearpage
\renewcommand{\thefootnote}{\alph{footnote}}
\begin{deluxetable}{llccc}
\tablecolumns{5}
\tablewidth{0pc}
\tabletypesize{\small}
\tablecaption{Spectroscopic [Fe/H] Data for the Literature Sample}
\tablehead{
\colhead{Name} & \colhead{Study} & \colhead{$\rm{N_{stars}}$} & \colhead{$R\sim \frac{\lambda}{\Delta\lambda}$} & \colhead{[Fe/H] Method\tablenotemark{a}}
}
\startdata
NGC 104 & \cite{Lane11} & 2241 & 10000 & MRS\\
NGC 6218 & \cite{Lane11} & 242 & 10000 & MRS\\
NGC 6656 & \cite{Lane11} & 345 & 10000 & MRS\\
NGC 5024 & \cite{Lane11} & 180 & 10000 & MRS\\
NGC 288 & \cite{Lane11} & 133 & 10000 & MRS\\
NGC 6752 & \cite{Lane11} & 437 & 10000  & MRS\\
NGC 1904 & \cite{CG09} & 58 & 40000 & HRS\\
NGC 3201 & \cite{CG09} & 149 & 40000 & HRS\\
NGC 4590 & \cite{CG09} & 122 & 40000 & HRS\\
NGC 5904 & \cite{CG09} & 136 & 40000 & HRS\\
NGC 6121 & \cite{CG09} & 103 & 40000 & HRS\\
NGC 6171 & \cite{CG09} & 33 & 40000 & HRS\\
NGC 6254 & \cite{CG09} & 147 & 40000 & HRS\\
NGC 6388 & \cite{CG09} & 36 & 40000 & HRS\\
NGC 6397 & \cite{CG09} & 144 & 40000 & HRS\\
NGC 6809 & \cite{CG09} & 157 & 40000 & HRS\\
NGC 6838 & \cite{CG09} & 39 & 40000 & HRS\\
NGC 7078 & \cite{CG09} & 85 & 40000 & HRS\\
NGC 7099 & \cite{CG09} & 65 & 40000 & HRS\\
M 54/Sgr & \cite{Carretta10} & 103 & 40000 & HRS\\
NGC 1851 & \cite{Carretta11} & 124 & 40000 & HRS\\
IC 4499 & \cite{Hankey11} & 43 & 10000 & CaT\\
Trumpler 5 & \cite{Cole04} & 15 & 3400 & CaT\\
NGC 6791 & \cite{Warren09} & 23 & 10000 & CaT\\
NGC 6819 & \cite{Warren09} & 9 & 10000 & CaT\\
NGC 6939 & \cite{Warren09} & 13 & 10000 & CaT\\
NGC 7142 & \cite{Warren09} & 6 & 10000 & CaT\\
NGC 7789 & \cite{Warren09} & 20 & 10000 & CaT\\
Berkeley 99 & \cite{Warren09} & 9 & 10000 & CaT\\
King 2 & \cite{Warren09} & 7 & 10000 & CaT\\
NGC 7044 & \cite{Warren09} & 10 & 10000 & CaT\\
Berkeley 39 & \cite{Warren09} & 10 & 10000 & CaT\\
M 67 & \cite{Warren09} & 7 & 10000 & CaT\\
Melotte 66 & \cite{Warren09} & 13 & 10000 & CaT\\
NGC 2141 & \cite{Warren09} & 14 & 10000 & CaT\\
NGC 2419 & \cite{Cohen11} & 7 & 34000 & HRS\\
NGC6205 & \cite{KI03} & 27 & 45000 & HRS\\
NGC 5272 & \cite{Sneden04} & 23 & 45000 & HRS\\
NGC 6341 & \cite{Sneden00} & 33 & 20000 & HRS\\
NGC 362 & \cite{Shetrone00} & 12 & 30000 & HRS\\
Terzan 5 & \cite{Origlia04} & 6 & 25000 & HRS\\
Pyxis & \cite{Saviane12} & 5 & 3400 & CaT\\
NGC 2808 & \cite{Saviane12} & 17 & 3400 & CaT\\
Rup 106 & \cite{Saviane12} & 9 & 3400 & CaT\\
NGC 5824 & \cite{Saviane12} & 17 & 3400 & CaT\\
Lynga 7 & \cite{Saviane12} & 8 & 3400 & CaT\\
NGC 6139 & \cite{Saviane12} & 15 & 3400 & CaT\\
Terzan 3 & \cite{Saviane12} & 13 & 3400 & CaT\\
NGC 6325 & \cite{Saviane12} & 10 & 3400 & CaT\\
NGC 6356 & \cite{Saviane12} & 11 & 3400 & CaT\\
HP 1 & \cite{Saviane12} & 8 & 3400 & CaT\\
NGC 6380 & \cite{Saviane12} & 8 & 3400 & CaT\\
NGC 6440 & \cite{Saviane12} & 8 & 3400 & CaT\\
NGC 6558 & \cite{Saviane12} & 4 & 3400 & CaT\\
Pal 7 & \cite{Saviane12} & 14 & 3400 & CaT\\
NGC 6569 & \cite{Saviane12} & 7 & 3400 & CaT\\
Terzan 7 & \cite{Saviane12} & 12 & 3400 & CaT\\
NGC 7006 & \cite{Saviane12} & 20 & 3400 & CaT\\
NGC 6441 & \cite{Saviane12} & 7 & 3400 & CaT\\
NGC 6528 & \cite{Saviane12} & 4 & 3400 & CaT\\
NGC 6553 & \cite{Saviane12} & 13 & 3400 & CaT\\
%
WLM & \cite{Leaman09,Leaman12} & 126 & 3400,7000 & CaT\\
LMC & \cite{Cole05,Carrera08b} & 815 & 3400 & CaT\\
SMC & \cite{Carrera08,Parisi10} & 729 & 3400 & CaT\\
Fornax & \cite{Battaglia06} & 870 & 6500 & CaT\\
Sculptor & \cite{Tolstoy04} & 629 & 6500 & CaT\\
Sextans & \cite{Battaglia11} & 180 & 6500 & CaT\\
Carina & \cite{Koch06} & 327 & 6500 & CaT\\
Leo I & \cite{Kirby10} & 827 & 7000 & MRS\\
Leo II & \cite{Kirby10} & 258 & 7000 & MRS\\
Canes Venitci I & \cite{Kirby10} & 170 & 7000 & MRS\\
Draco & \cite{Kirby10} & 298 & 7000 & MRS\\
Ursa Minor & \cite{Kirby10} & 212 & 7000 & MRS\\
Hercules & \cite{Aden11} & 11 & 20000 & HRS\\
$\omega$Cen & \cite{Stanford06} & 378 & 2000 & MRS\\
 & \cite{Johnson08,Johnson09} & 246 & 13000,18000 & HRS\\
Willman I & \cite{Martin07} & 8 & 7000 & CaT\\
Ursa Major I & \cite{Martin07} & 10 & 7000 & CaT\\
Tucana & \cite{Fraternali09} & 14 & 3400 & CaT\\
NGC 6822 & \cite{Tolstoy01} & 26 & 3400 & CaT\\
Segue I & \cite{Norris10,Geha09} & 4 & 5000,7000 & MRS\\
Bootes I & \cite{Lai11} & 25 & 2000 & MRS\\
\enddata
\tablenotetext{a}{Method which the the original study used to compute [Fe/H] estimates: Spectrum synthesis based techniques on medium (MRS) or high resolution spectra (HRS); empirical calibration based on the Calcium II Triplet (CaT) features.}
\end{deluxetable}
\renewcommand{\thefootnote}{\arabic{footnote}}

\clearpage
\renewcommand{\thefootnote}{\alph{footnote}}
\begin{deluxetable}{lcccc}
\tablecolumns{5}
\tablewidth{0pc}
\tabletypesize{\small}
\tablecaption{Intrinsic Metallicity Spreads}
\tablehead{
\colhead{Name} & \colhead{$\overline{[Fe/H]}$} & \colhead{$\sigma([Fe/H])$\tablenotemark{a}} & \colhead{$\bar{Z}$} & \colhead{$\sigma(Z)^{2}$\tablenotemark{a}}
}
\startdata
    NGC 104 & -0.82 &  0.023 & 1.73E-01 & 1.02E-03\\
   NGC 6218 & -1.15 &  0.061 & 8.04E-02 & 2.84E-04\\
   NGC 6656 & -1.95 & $\leq$0.197 & 1.26E-02 & 4.09E-06\\
   NGC 5024 & -1.89 &  0.040 & 1.44E-02 & 9.26E-06\\
   NGC 288 & -1.25 & $\leq$0.195 & 6.22E-02 & 1.39E-04\\
   NGC 6752 & -1.45 & $\leq$0.145 & 3.74E-02 & 1.57E-04\\
   NGC 1904 & -1.54 & $\leq$0.035 & 2.89E-02 & $\leq$5.52E-06\\
   NGC 3201 & -1.49 &  0.048 & 3.25E-02 & 1.32E-05\\
   NGC 4590 & -2.22 &  0.076 & 6.12E-03 & 1.29E-08\\
   NGC 5904 & -1.35 & $\leq$0.024 & 4.52E-02 & $\leq$6.13E-06\\
   NGC 6121 & -1.20 & $\leq$0.025 & 6.33E-02 & $\leq$1.30E-05\\
   NGC 6171 & -1.06 & $\leq$0.043 & 8.67E-02 & $\leq$7.11E-05\\
   NGC 6254 & -1.56 &  0.052 & 2.80E-02 & 1.17E-05\\
   NGC 6388 & -0.40 &  0.074 & 4.04E-01 & 4.38E-03\\
   NGC 6397 & -1.99 &  0.003 & 1.02E-02 & 4.04E-09\\
   NGC 6809 & -1.97 &  0.044 & 1.09E-02 & 1.20E-06\\
   NGC 6838 & -0.81 & $\leq$0.033 & 1.56E-01 & $\leq$1.34E-04\\
   NGC 7078 & -2.34 &  0.011 & 4.61E-03 & 2.59E-08\\
   NGC 7099 & -2.36 &  0.005 & 4.42E-03 & 3.34E-09\\
   M54/Sgr\tablenotemark{b} & -1.29 &  0.445 & 1.09E-01 & 1.18E-02\\
   NGC 1851 & -1.14 &  0.058 & 7.27E-02 & 9.74E-05\\
   NGC 6205 & -1.61 & $\leq$0.055 & 2.45E-02 & $\leq$9.26E-06\\
  IC 4499 & -1.15 & $\leq$0.107 & 7.33E-02 & $\leq$3.56E-04\\
     Trumpler 5 & -0.49 &  0.124 & 3.60E-01 & 7.76E-03\\
   NGC 6791 &  0.39 &  0.147 & 2.68E+00 & 5.75E-01\\
   NGC 6819 & -0.06 & $\leq$0.108 & 9.07E-01 & 6.32E-02\\
   NGC 6939 & -0.22 & $\leq$0.039 & 6.04E-01 & $\leq$3.18E-03\\
   NGC 7142 & -0.19 & $\leq$0.139 & 6.65E-01 & 3.84E-02\\
   NGC 7789 & -0.23 & $\leq$0.103 & 6.13E-01 & 3.40E-02\\
    Berkeley 99 & -0.57 &  0.034 & 2.97E-01 & 4.42E-03\\
   King 2 & -0.42 & $\leq$0.133 & 3.99E-01 & $\leq$1.33E-02\\
   NGC 7044 & -0.15 &  0.181 & 7.57E-01 & 1.51E-02\\
    Berkeley 39 & -0.38 & $\leq$0.066 & 4.20E-01 & $\leq$4.03E-03\\
     M 67 & -0.19 & $\leq$0.042 & 6.54E-01 & $\leq$3.92E-03\\
   Melotte 66 & -0.49 & $\leq$0.049 & 3.26E-01 & $\leq$1.32E-03\\
   NGC 2141 & -0.27 & $\leq$0.085 & 5.46E-01 & 1.02E-02\\
   NGC 2419 & -2.06 & $\leq$0.096 & 8.86E-03 & 3.50E-06\\
    NGC 362 & -1.33 & $\leq$0.034 & 4.74E-02 & $\leq$1.32E-05\\
   NGC 5272 & -1.58 & $\leq$0.069 & 2.65E-02 & $\leq$1.49E-05\\
   NGC 6341 & -2.34 & $\leq$0.039 & 4.58E-03 & $\leq$1.49E-07\\
    Ter 5 & -0.25 & $\leq$0.057 & 5.73E-01 & $\leq$5.20E-03\\
     WLM & -1.28 &  0.387 & 8.26E-02 & 1.62E-03\\
     LMC & -0.55 &  0.367 & 3.83E-01 & 3.14E-02\\
     SMC & -1.07 &  0.349 & 1.46E-01 & 5.65E-03\\
     Fornax & -1.29 &  0.446 & 8.40E-02 & 2.43E-03\\
     Sculptor & -1.95 &  0.355 & 1.77E-02 & 1.57E-04\\
     Sextans & -2.29 &  0.387 & 1.17E-02 & 5.74E-05\\
     Carina & -1.94 &  0.346 & 1.68E-02 & 1.05E-04\\
    Leo I & -1.46 &  0.286 & 4.33E-02 & 3.88E-04\\
    Leo II & -1.70 &  0.362 & 2.79E-02 & 2.22E-04\\
    Canes Venitici I & -2.01 &  0.446 & 2.03E-02 & 1.69E-04\\
    Bootes I & -2.59 &  0.373 & 4.00E-03 & 3.89E-06\\
    Segue I & -2.73 &  0.626 & 4.93E-03 & 1.15E-05\\
     Tucana & -1.97 &  0.265 & 1.34E-02 & 3.90E-05\\
   NGC 6822 & -1.20 &  0.457 & 1.18E-01 & 3.78E-03\\
    Ursa Major I & -2.07 &  0.293 & 1.24E-02 & 4.62E-05\\
    Willman I & -1.40 &  0.324 & 5.62E-02 & 8.11E-04\\
     Draco & -1.99 &  0.363 & 1.72E-02 & 9.45E-05\\
     Ursa Minor & -2.13 &  0.339 & 1.40E-02 & 6.12E-05\\
    Hercules & -2.70 &  0.331 & 2.81E-03 & 2.83E-06\\
    $\omega$Cen & -1.51 &  0.253 & 3.93E-02 & 6.95E-04\\
   Pyxis & -1.09 &  0.246 & 9.27E-02 & 1.57E-05\\
   NGC 2808 & -0.87 &  0.106 & 1.40E-01 & 7.61E-05\\
  Rup 106 & -1.46 &  0.062 & 3.61E-02 & 2.52E-05\\
   NGC 5824 & -1.74 &  0.115 & 1.93E-02 & 2.41E-05\\
  Lynga 7 & -0.57 & $\leq$0.035 & 2.72E-01 & $\leq$4.59E-04\\
   NGC 6139 & -1.32 &  0.069 & 4.88E-02 & 5.50E-05\\
    Ter 3 & -0.81 &  0.070 & 1.62E-01 & 4.59E-04\\
   NGC 6325 & -1.10 &  0.012 & 8.06E-02 & 2.75E-05\\
   NGC 6356 & -0.51 &  0.061 & 3.15E-01 & 2.18E-03\\
    HP 1 & -1.04 &  0.084 & 9.53E-02 & 3.10E-04\\
   NGC 6380 & -0.51 &  0.073 & 3.13E-01 & 2.48E-03\\
   NGC 6440 & -0.39 &  0.077 & 4.10E-01 & 4.94E-03\\
   NGC 6558 & -0.83 & $\leq$0.069 & 1.48E-01 & 5.72E-04\\
   Pal 7 & -0.49 & $\leq$0.082 & 3.29E-01 & $\leq$3.60E-03\\
   NGC 6569 & -0.82 & $\leq$0.088 & 1.53E-01 & 1.12E-03\\
   Terzan 7 & -0.32 &  0.089 & 4.93E-01 & 1.20E-02\\
   NGC 7006 & -1.32 &  0.074 & 5.04E-02 & 6.71E-05\\
   NGC 6441 & -0.53 & $\leq$0.080 & 3.03E-01 & $\leq$3.23E-03\\
   NGC 6528 & -0.44 & $\leq$0.076 & 3.64E-01 & $\leq$3.67E-03\\
   NGC 6553 & -0.28 & $\leq$0.082 & 5.34E-01 & $\leq$9.43E-03\\
\enddata
\tablenotetext{a}{Column reports intrinsic spreads - observational measurement errors have been removed following Equation 1}
\tablenotetext{b}{We consider all stars in the M54 and Sgr region from this sample and note that due to the extreme difficulty in associating stars simply to M54 or Sgr (see \citealt{Carretta10}), the values in this table should be considered upper limits}
\tablecomments{Values of upper limits ($\leq$) reported where intrinsic spread is smaller than the lowest measurement error of an individual star from the original study (see $\S4$).} 
\end{deluxetable}
\renewcommand{\thefootnote}{\arabic{footnote}}
%
%
%
%
%


\end{document}